# The Way To Circumbinary Planets

Hans J. Deeg[1,2] and Laurance R. Doyle[3]

[1]Instituto de Astrofísica de Canarias, C/ Via Lactea s/n, 38205 La Laguna, Tenerife, Spain

[2]Universidad de La Laguna, Dept. de Astrofísica, 38206 La Laguna, Tenerife, Spain

hdeeg@iac.es

[3]Carl Sagan Center, SETI Institute, 339 Bernardo Avenue, Mountain View, California 94043

# Abstract

Circumbinary planets (CBPs) are planets that orbit around both stars of a binary system. This chapter traces the history of research on CBPs and provides an overview over the current knowledge about CBPs and their detection methods. After early speculations about CBPs, inspired by binary star systems and popularized by fictional works, their scientific exploration began with the identification of circumbinary dust disks and progressed to the detection and characterization of the current sample of CBPs. The major part of this review presents the detection methods for CBPs: eclipse timing variations from the light-travel-time effect and from dynamical interactions, transits, radial velocities, direct imaging, gravitational microlensing and astrometry. Each of these methods is described with its strengths and limitations and the main characeristics of the CBP systems found by them are outlined. The potential habitability of CBPs is considered, taking into account the unique environmental conditions created by orbiting a stellar binary. The importance of multi-method detection strategies is underscored, and future advancements from upcoming missions like PLATO are anticipated, promising to expand the understanding of these intriguing celestial bodies.

# Historical Introduction

The nature of eclipsing binary stars was first correctly explained in 1783 by the Dutch astronomer John Goodricke, while working in England. He proposed that variable stars like Algol were actually binary systems whose orbital plane caused them to regularly eclipse each other across our line-of-sight (Goodricke 1784). Also, long before the discovery of the first extrasolar planets, speculations were made about the existence and nature of planets in or around binary systems. These were largely driven by the strangeness of such imaginary planets, with two suns being visible from their surface, each with its own morning and evening (Fig. 1; see also Fig. 10). Science fiction provided certainly the most famous such planet, where the desert planet Tatooine, circled by two solar-like suns, provided the opening setting for George Lucas' original 1977 Star Wars movie, with several more appearances in later sequels of that space opera. Further examples of CBPs in fiction can be found in the Wikipedia entry for circumbinary planets.

Closer to the scientific realm, the first documented remarks about planets in binary systems are likely those from Camille Flammarion (1874), who wrote "Les étoiles doubles sont donc en réalité des groupes de deux soleils. Ces soleils gravitent l'un autour de l'autre, et il est bien probable, pour ne pas dire certain, qu'autour de chacun de ces foyers une famille de planètes est suspendue" ("Double stars are in fact groups of two suns. These suns revolve around each other, and it is quite likely, not to say certain, that around each of them there is a system of planets"). While this seems to refer to planets around each component of a binary, Flammarion might have had circumbinary planets in mind when he wrote a few years later (Flammarion 1884): "Quelles merveilleuses années, quelles singulières saisons, quels jours et quelles nuits fantastiques sont le partage des planètes inconnues qui gravitent autour de ces [deux] étoiles colorées" ("What wonderful years, what singular seasons, what days and what fantastic nights are sharing unknown planets which revolve around these [two] colored stars").

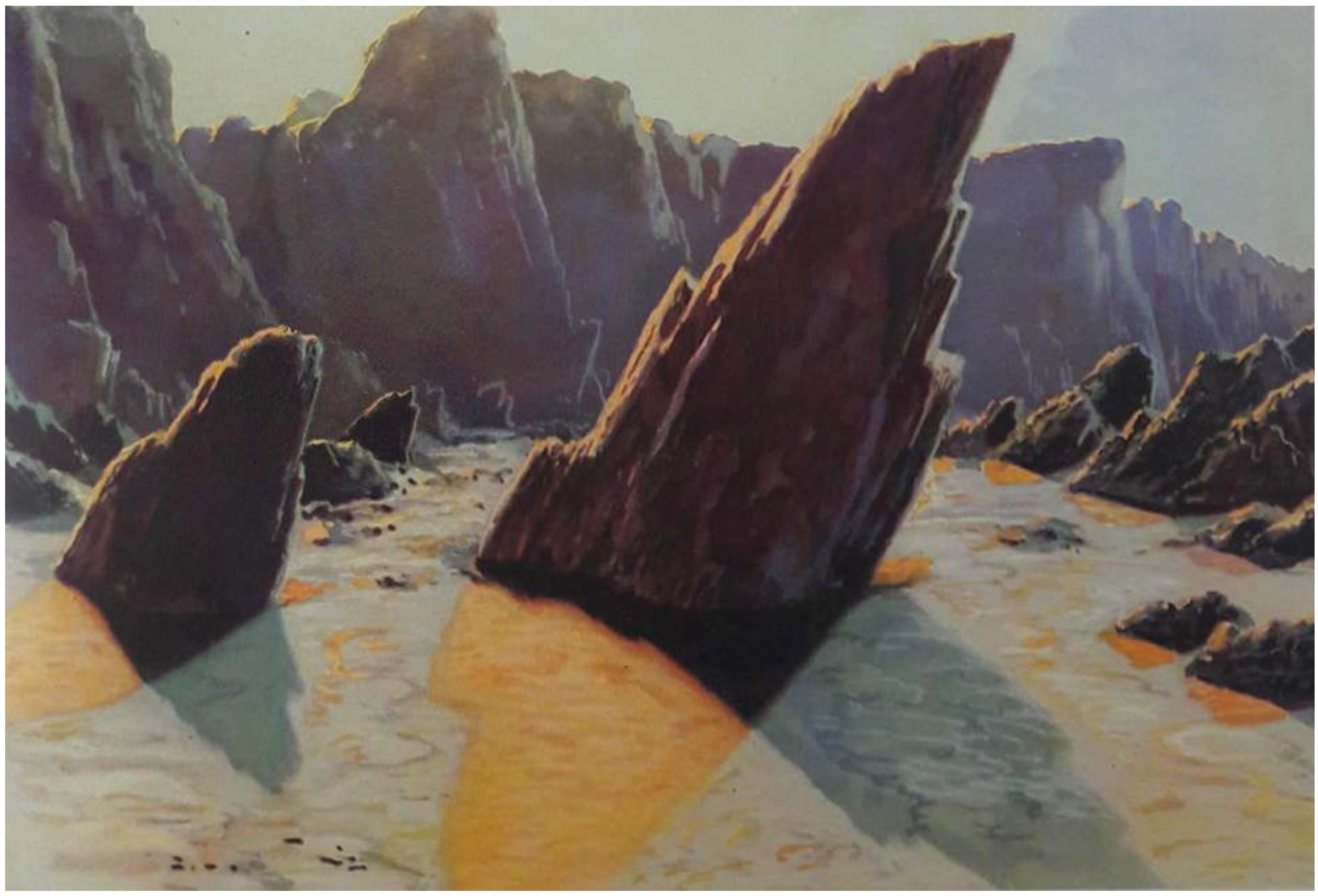

Fig. 1. Illustration from the book "Sur les autres mondes" (Rudaux 1937), with a planet surface illuminated by two suns of different colors. While the shadows' colors might be realistic, modern stability criteria of circumbinary planet orbits imply that the angle between the shadows is too large, requiring a planet that is too close to the binary for a stable orbit. Credit: Photographic reproduction from original book by Lucien Rudaux (1874-1947), in public domain.

Arriving at a modern scientific viewpoint, the existence of CBPs was not entirely unexpected, due to the discovery of circumbinary (potentially protoplanetary) dust-disks. For short-periodic evolved binary systems, evidence for such disks had been accumulating since the 1970s (e.g., the V471 Tau system,

Paczynski 1976), with some systems also giving direct observational evidence from spectroscopy (e.g. epsilon Aurigae, Castelli 1977; Beta Lyrae, Kondo et al. 1983). Direct evidence for circumbinary envelopes was later detected in images taken by the Hubble Space Telescope (HST) -- the circumbinary disc around the GG Tau system being the best example (Krist et al., 2002, McCabe et al. 2002; see also Fig. 2). Although it was argued much later (Nelson & Marzari 2016) that GG Tau's disk is not a suitable place for planet formation, these images certainly raised the expectation that CBPs are able to form and be detected.

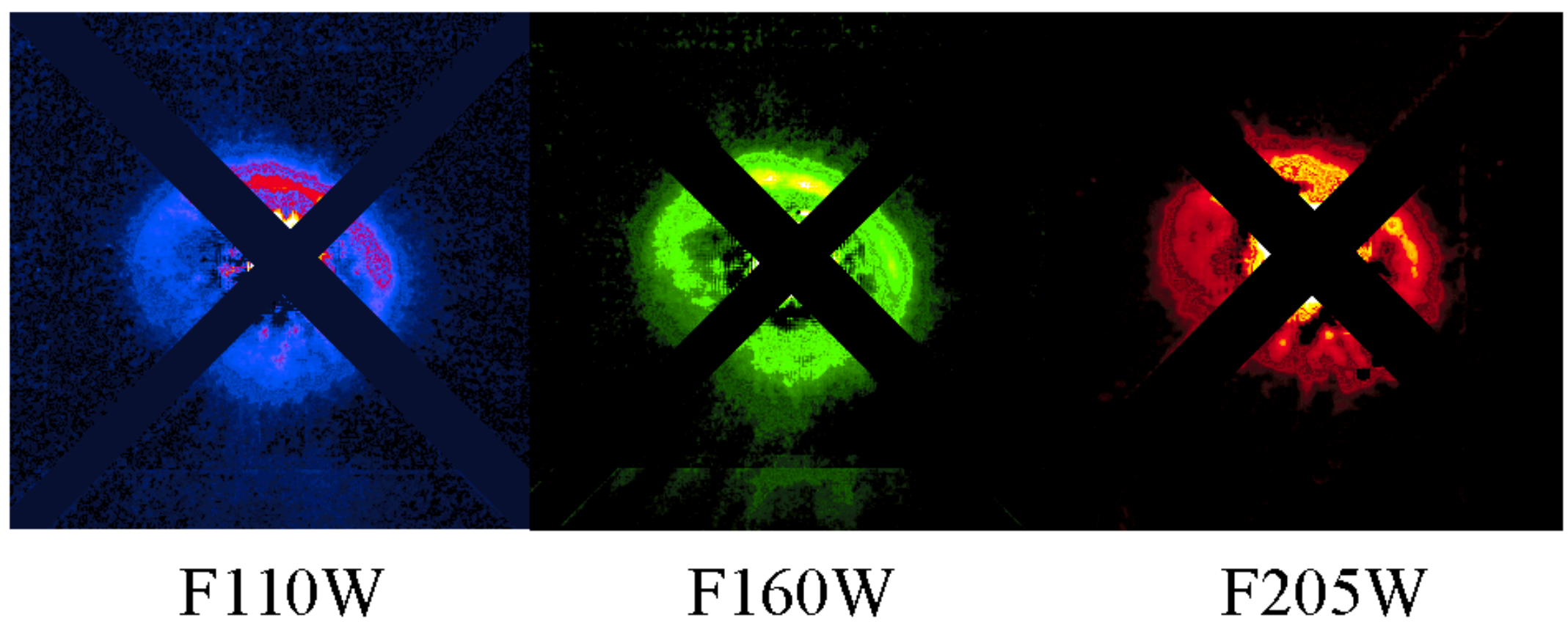

Fig. 2. The circumbinary ring around the binary system GG Tau, after removal of the light from the central binary; in three HST-NICMOS bands at 1.1, 1.6 and 2.05 µm (based on McCabe et al. 2002, reproduced with permission of the authors).

Obviously, it only made sense to speculate seriously about the existence of CBPs if it could be demonstrated that such planets could accrete from circumbinary discs and achieve stable orbits. The first investigations to directly focus on CBP formation were indeed investigations into their orbital stability – a field that had evolved from more general considerations of the orbital stability of the 3-body problem (e.g. Black 1982, Pendleton & Black 1983). The pioneering work by Rudolf Dvorak (1986) introduced the still-used nomenclature of 'P-type' (Planet-type) orbits for CBPs, and 'S-type' (Satellite-type) orbits for objects orbiting only one of the components in a wide binary (the "circumstellar" planets). In this paper, and in Dvorak (1989), he established that CBPs can only have a stable orbit if their distance from the common 3-body barycenter is $\gtrsim$ 2.3 times the distance between two binary components (for circular orbits), which corresponds to a planet to binary period ratio of $\gtrsim$ 3.5. For high eccentricities, this distance ratio increases to about 4, whereas it has only a slight dependence on the mass-ratio between the binary components. This stability criterium also implies that the angle between the 'two suns', as seen from the surface of a CBP, cannot exceed 24° (hence the angle between shadows in the historic illustration of Fig. 1 is too large, although projection effects may increase the angle between shadows against the angle between the binary components, but not as much as in that figure). To this day, the work of reference on orbital stability of both P and S-type planets is the one by Holman & Wiegert (1999), with detailed stability criteria for different orbital eccentricities and mass-ratios of the stellar components. The stability of 'misaligned' CBPs with high orbital inclinations (up to 50°) against the binary plane was investigated by Pilat-Lohinger et al. (2003), who found that inclination has no significant influence on orbital stability. Morais & Giuppone (2012) considered the orbital stability of both prograde and retrograde planets. Doolin & Blundell (2011) focused on long-term dynamics (libration, precession) and stability of CBP orbits, while Chavez et al. (2014) investigated the long-term stability of some of the CBPs found by the Kepler mission. (See also the chapter Populations of planets in binary star systems for a more thorough discussion of CBP stability)

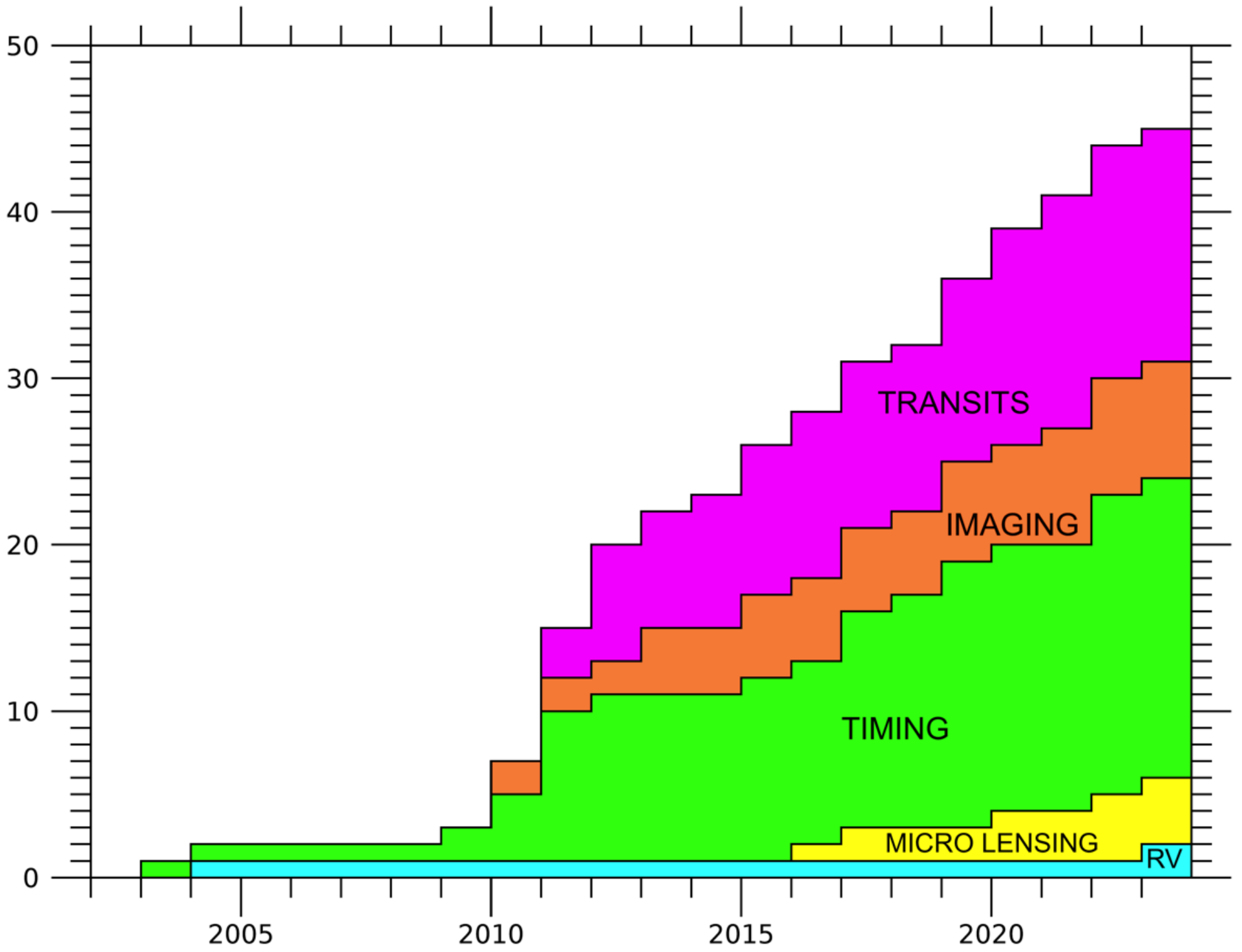

Fig. 3. Cumulative count of CBPs, against the year of their publication. Based on data from the NASA Exoplanet Archive (2024), using its classification by discovery methods. 'RV' means radial velocities. 'Timing' includes the first of all CBPs, found by pulsar timing (PSR B1620-26 b, accepted as a CBP in 2003) and the one planet found by dynamical interactions (Kepler-1660AB b), while all other 'timing' planets were found by eclipse time variations due to the light-time travel effect (LTTE).

Similar to the discovery of the first exoplanets around single stars, which were found by the timing of pulsar signals (see chapter Pulsar Timing as an Exoplanet Discovery Method), the first circumbinary planet was discovered by the timing of such signals, in this case around a pulsar-white dwarf binary (Backer et al. 1993). Since then, a current total of 45 CBPs have been discovered by a variety of detection methods (from NASA Exoplanet Archive as of April 2024; Fig. 3), which are outlined in the remainder of this chapter. These 45 CBPs are around 36 different binaries, with seven binaries that host more than one known CBP (5 systems with 2 CBPs and 2 systems with 3 CBPs). These methods predominantly select CBPs in some limited regions in parameter space, as shown by Fig. 4, when plotting them against the binaries' and CBPs' orbital periods. These regions are consequences of both the different sensitivities of the various methods, but also of the underlying intrinsic CBP distributions, which are still in an early phase to be revealed. It is of note that no consistent naming scheme of CBPs has evolved. All transiting CBPs that are first-discovered planets around a given binary are designated with a 'b', similar to planets around single stars. However, some CBPs discovered by other methods come with a 'c' (without the existence of a planet 'b') ; examples are NN Ser c, Ross 458 c, HD 202206 c, and OGLE-2019-BLG-1470L AB c, discovered by eclipse timing, imaging, radial velocities and microlensing, respectively. In these cases, the 'c' was assigned in continuation of the sequence of the stellar components A and B (see chapter The Naming of Extrasolar Planets).

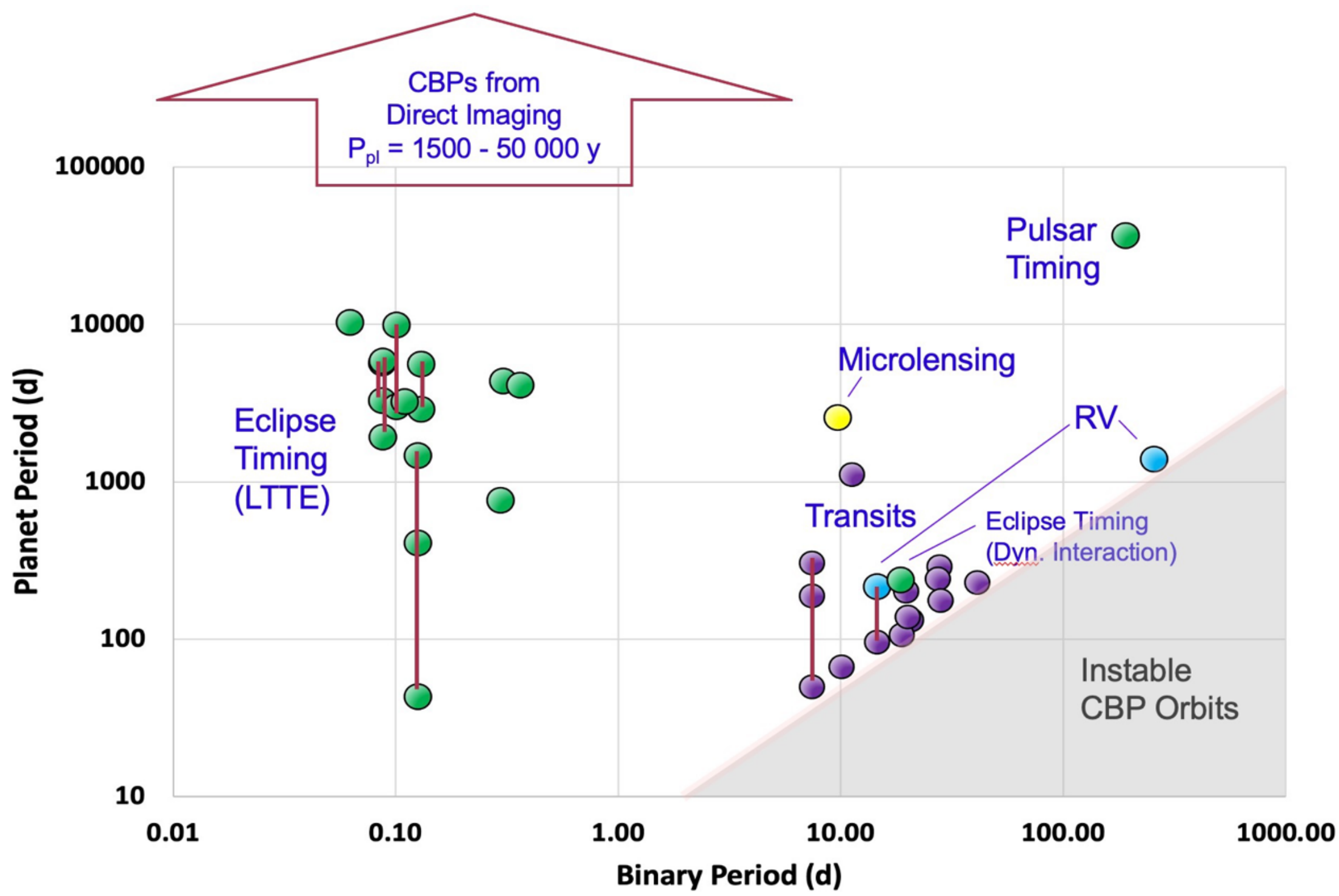

Figure 4. The currently known CBPs, plotted by planetary orbital period versus the orbital period of their central binary. The detection methods are indicated by colors that are similar to Fig. 3. Planets in multi-planet systems are indicated by connection-lines. The seven CBPs detected by imaging have very long periods with large uncertainties and are outside of the plotted period range. Also, only one of the five planets from micro-lensing has a reliable period value; the others aren't shown. Data from the NASA Exoplanet Archive (2024), selecting all planets with a circumbinary flag of 1. The zone of instable orbits corresponds to planetary orbital periods that are less than 4 times the binary period.

# Circumbinary Planet Detection Methods

In the following, the methods that have led to successful detections or characterization of CBPs are presented first, and then the methods that are still to be proven for these tasks.

## Eclipse Timing Variations from the Light-Travel-Time Effect (LTTE)

Over a century ago, researchers working on short-periodic binaries (Woltjer 1922, for the interpretation of eclipses of RZ Cas; see also Irwin 1952, 1959) deduced that orbiting third bodies could be responsible for slight variations in the periodicity of stellar binary eclipse minima times (known as eclipse time variations or ETVs). One potential cause of ETVs, the Rømer effect or light-travel-time effect (LTTE) had been known since the late 17$^{th}$ century from observations of occultations of Jupiter's satellite Io by Jupiter itself, when Rømer (1677) measured a deviation of occultation times from strict periodicity. He correctly interpreted this to being due to the finite speed of light and derived a first estimate for it (also described in Huygens 1677).

In general terms, the LTTE is caused by a variation in the distance between an observer and an astronomical object that emits a periodic signal. This object might be a binary with its mutual eclipses (which provide observable ETVs), but it could also be the light-pulses from pulsars (see chapter PSR B1257+12 and the First Confirmed Planets Beyond the Solar System), or stars with pulsations that are known to have very stable intrinsic periods. An example are hot subdwarf B stars (spectral type sdB) that usually display very constant pulsations, for which some cases with deviations from strict periodicities were initially interpreted as LTTEs due to orbiting planets (e.g. Silvotti et al. 2007 for V391 Peg , Charpinet et al 2011 for Kepler 70); albeit these were later superseded by alternative explications, (e.g. Silvotti et al. 2018, Blokesz et al. 2019).

In the case when a further body orbits a periodic emitter, the emitter will be displaced around the common barycenter by some distance. If the emitter is an eclipsing binary, interpretations based on the presence of an LTTE assume that the binary orbital parameters are unperturbed by the third body, except for a displacement of the entire binary around the barycenter (if orbital parameters are perturbed, this is considered a dynamical interaction, discussed further below). The observed distance from Earth to the emitter then varies with the orbital period of the further body, and light from the emitter needs to travel varying distances until its reception on Earth. The amplitude $K_{\mathrm{O\text{-}C}}$ of the timing-variation (the maximum temporal advance or delay of observed eclipse events versus ones of strict periodicity) is given by (e.g. Deeg et al 2008):

$$K_{\mathrm{O-C}} = \frac{m_3}{(m_{Bin})^{2/3}\, c} \left(\frac{P_3^2\, G}{4\,\pi^2}\right)^{1/3} \sin i_3 \qquad (1)$$

Where $m_3$, $P_3$ and $i_3$ is the mass, orbital period and orbital inclination of the orbiting body (if the emitter is an EB, the outer body is referred to as 'third body'), $m_{Bin}$ is the total mass of the periodic emitter (in the case of a binary, its total mass), *c* the speed of light, and *G* the gravitational constant. It is clear that this method works best for massive and long-periodic orbiting bodies, and it is also most effective for low-mass mass emitters (see Sybilski et al. 2010 for a detailed investigation of the LTTE 'discovery space' under the performance parameters of the CoRoT and Kepler space missions). Before interpreting any variations in periodicity as an LTTE, they need to be corrected for the yearly motion of Earth around the Earth-Sun barycenter. The times at which an astronomical event is observed on Earth (geocentric dates) are therefore converted to heliocentric (HJD) or solar-system barycentric Julian dates (BJD), with BJD being required for the timing of events that need precisions on the order of seconds or less. For an excellent overview over the requirements to achieve such precisions and the different time standards, see Eastman et al. (2010).

Only with the development of modern photometric techniques for CCDs and the wide availability of precise time-standards (initially time signals emitted by radio stations, and now spaced based services like GPS) did observations become sufficiently precise and well-calibrated to attempt detections by LTTE of CBPs of detached MS-EBs (main-sequence EBs). LTTE detections are somewhat easier from the eclipse timing of binaries made of compact objects – such as eclipses by white dwarfs – or from very short-periodic binaries, like the post common envelope binaries (PCEBs). This is mainly due to the intrinsic relation between the precision of an eclipse minimum time measurement against the photometric precision (the precision of the flux-measurement), the relative depth of the eclipses, and the duration of the eclipses' in- and egress. This relation can be approximated by (Deeg & Tingley 2017):

$$\sigma_{\mathrm{t}} = \frac{\sigma_{\mathrm{F}\tau}}{2\,\Delta\mathrm{F}} \sqrt{\tau\, \mathrm{T}_{\nabla}} \qquad (2)$$

Where $\sigma_t$ is the timing precision, $\sigma_{F\tau}$ is the photometric precision of a time-series over a time-scale $\tau$, $\Delta F$ is the depth of the eclipse, and $T_{\nabla}$ is the combined duration of the in- and egress, excluding the flat central part of an eclipse (if there is any). Timings of very short eclipses among compact objects like white dwarfs are therefore measurable with a much higher precision than those from MS-EBs, and these also led to the first claims of CBP detections. A candidate CBP around the pulsar-white dwarf binary PSR B-1620-26 in the globular cluster M4 was proposed as one of two possibilities in 1993 (Thorsett et al. 1993, Backer 1993). Secondary variations in the pulsar timings indicated that the system was orbited either by a third star with a semi-major axis of 50 AU, or by a CBP with a semi-major axis of 10 AU. It took ten more years before the orbiting third body was reliably characterized and found to be a 2.5 jovian-mass CBP, with a semi-major axis of 23 AU (Sigurdsson et al. 2003). Hence, it was not until the 21st century that the first claims for the detection of CBPs were firmly established.

The confirmation of the presence of an LTTE is not a trivial task. In the case of EBs, deviations from periodicity may also be due to intrinsic variations in the binary period due to a variety of effects: mass-transfers between stellar components in the case of contact or semi-contact EB systems; emission of strong winds from stellar surfaces; variations in the shape of a magnetically active binary components, known as the Applegate effect (Applegate 1992); starspot activity which modifies the shapes of binary eclipses and then distorts eclipse minimum timing measurements (Watson & Dhillon 2004); and dynamic gravitational interactions of the third body with individual binary components. Dynamical gravitational interactions are known to modify eclipse times of several CBP systems discovered by transits, with much larger effects onto the eclipse timings than the LTTE (see later in this chapter). In any case, claims for CBP detections by LTTE need a careful revision for the presence of any alternative effects. The true origin of a potential CPB based on the LTTE may in some cases also be resolved from astrometric measurements, as has been done recently for HW Vir, based on data from Gaia and Hipparchos (Baycroft et al. 2023).

Besides the PSR B-1620-26 system mentioned above, all other claims for CBP detections based on the LTTE are for planets with orbital periods ranging from year-long to thousands of years (with one exception, Kepler-451 d with P= 43 days, Esmer et al. 2022), that orbit short-periodic (P < 0.4 d) evolved binary systems. CBPs around these evolved binaries are likely a fundamentally different planet population than CBPs around MS binaries, and they may have formed from masses that were ejected from the central binaries during their post MS evolution. These systems and the theories for their formation are reviewed in more detail in the chapter **Circumbinary Planets Around Evolved Stars**.

Regarding CBPs around MS binaries, studies searching for them with the LTTE were proposed by Doyle et al. (1998), but they have only been able to place approximate or upper limits for their masses (e.g. for a hypothesized CBP around the M4 /M4 binary CM Dra, Deeg et al. 2000, 2008; see Fig. 5) or be able to derive limits to their abundances. Borkovits et al. (2016) present a comprehensive study of ETVs of 2600 binaries observed by the Kepler mission, extending on a previous study by Conroy et al. (2014): Around MS binaries, no CBPs are detected from the LTTE. However, they find over 200 stellar companions, indicating that a very large fraction of all binaries are part of triples; while they also note a deficit of triple systems with binary periods $\lesssim$1 d and outer periods between $\sim$50 and 200 d; this might correspond to the known deficit of transiting CBPs around short-period binaries (see the section on transits in this chapter).

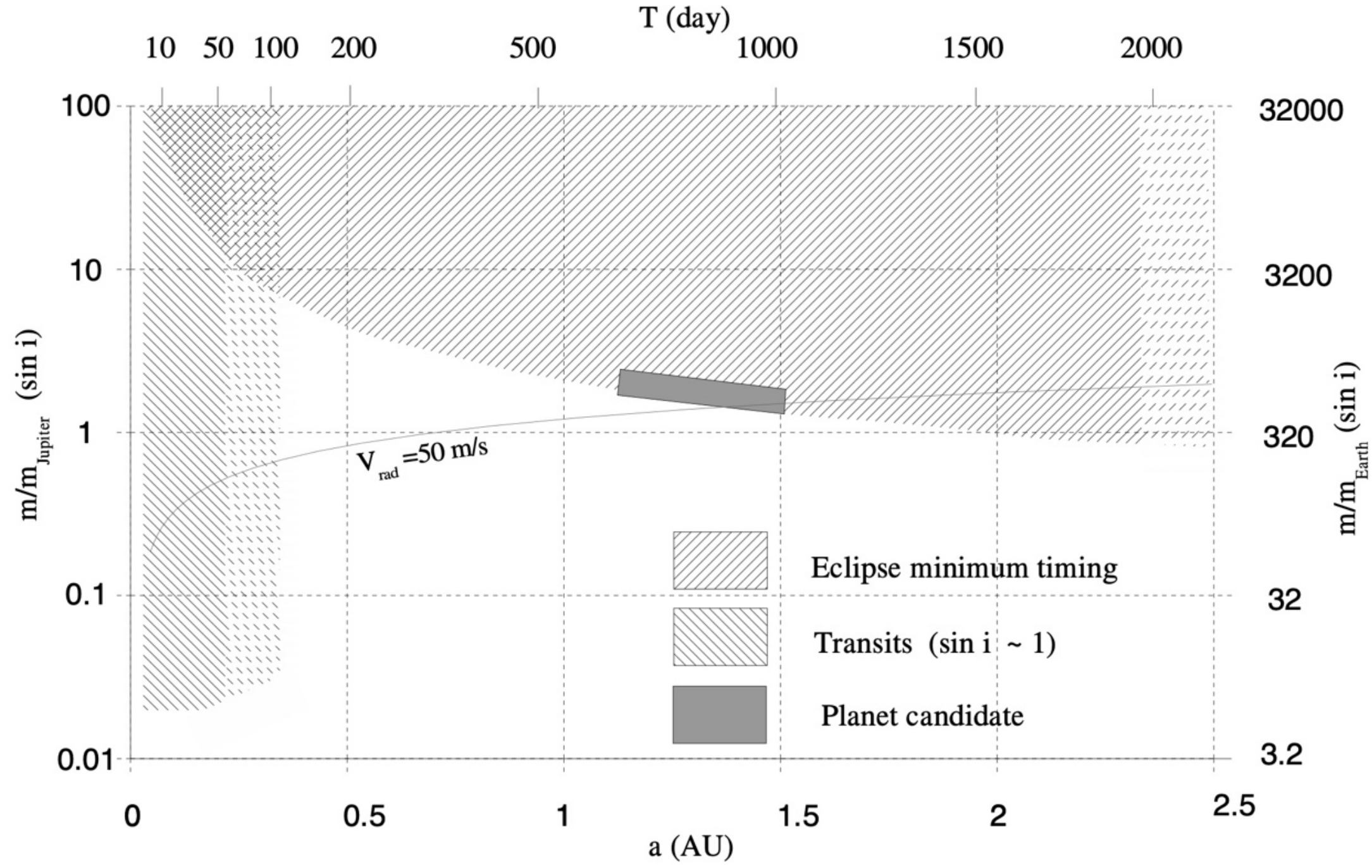

Fig. 5. Exclusion zones (hatched regions) for the detection of a CBP around the M4/M4 binary CM Dra. The horizontal axis gives the orbital distance of an assumed CBP from the binary's barycenter and the vertical one gives the CBP's mass. The hatched region on the left indicates that no transiting CBPs (of sizes $\gtrsim$ 2.5 $R_{Earth}$) are present at short orbital distances, while the upper hatched region indicates the absence of a planet with more than a few Jupiter masses, due to non-detection of an LTTE in eclipse timings. The dark grey zone outlines a weak LTTE detection of a candidate CBP. In the remaining white region, the detection methods had insufficient sensitivity. The black line gives the mass of a CBP with a radial velocity amplitude of 50 m/s. From Deeg et al. (2000); the candidate CBP could not be verified with later eclipse timings (Deeg et al. 2008).

## Transits by circumbinary planets

The idea that a planet around a single star could be orbiting through the line-of-sight to its central star, and the eclipses (usually called 'transits' when eclipses of planets across stars are meant) be photometrically detected, was initially suggested by the German astronomer Otto Struve (1952), with a first quantitative treatment by Rosenblatt (1971); see also the chapter Transit Photometry as an Exoplanet Discovery Method for an introduction to transits across single stars. The idea that this technique could be combined with the a-priori knowledge that EBs have edge-on orbits, making them preferential targets for transit-searches, originated with Borucki and Summers (1984) and, in more detail, with Schneider and Chevreton (1990). They argued that the orientation of EBs' orbits optimizes the likelihood that CBP orbital planes are also crossing our line-of-sight, citing models of protoplanetary discs, and the close alignment of the solar equator with the orbital planes of the solar system planets. Contrary to transits across single stars, transits from CBPs were also predicted (Deeg et al. 1998, Brandmeier & Doyle 1996) to produce brightness variations with irregular shapes that depend principally on the phase of the central binary during a planet's transit. CBP transits are semi-periodic, but always occur within a 'transit-window' given by the CBP's period (for details see Armstrong et al.

2011). The shapes and the timing of CBP transits are unique and permit unequivocal planet detection, without the need for confirmation by complementary observations, as is generally the case for transit detections around single stars.

Furthermore, Schneider (1994) pointed out that a CBP on an orbital plane with a relevant inclination against the binary plane is subject to precession, with rotating nodal lines between the two orbital planes. Depending on the orientation of the two intersecting planes, observable transits may or may not occur. For a CBP system status in which observable transits appear, albeit not necessarily at every CBP orbit, Martin & Triaud (2014) coined the term 'transitability'. Depending on a system's orientation and mutual binary-planet inclination, transitability of a given CBP may be permanent; or it may appear and disappear on decade-long precession time-scales; or it may never occur (transits will never be observable). For a more detailed discussion of CBP transits on mutually inclined orbits, including the case of CBP transits across non-eclipsing binaries, see also Martin & Triaud (2015) and Martin (2017).

Schneider and Doyle (1995) proposed that the EBs with the smallest stellar components – like the CM Draconis eclipsing binary system of two M4.5 components (with a combined surface area of ~12% of the solar disc) – would optimize any planetary transit signals, as the transit depth depends upon the ratio of the cross-section of the planet versus the combined cross-section of the binary components (note here that CBP transits during a binary eclipse are deeper than off-eclipse, since an EB's cross-section during eclipse is smaller). Also, CM Dra is known to have a near edge-on inclination, and such a low-mass system might furthermore permit the detection of more massive CBPs by eclipse timing from the LTTE (Deeg et al. 2000, 2008, Morales et al. 2009). Thus, the fairly bright (V=12.9, R = 10.8 mag) CM Draconis system became the primary target for TEP (transit of extrasolar planets). Between 1993 and 1999, this was the first observational search for exoplanet transits, operating as a network of observatories at different longitudes (Deeg et al. 1998). A CBP transit detection algorithm was developed for TEP (Doyle et al. 2000, based on Jenkins et al. 1996), which cross-correlated the observational light curves with all reasonable planetary transit sizes (i.e., planetary radii), and orbital characteristics (i.e., period, inclination, eccentricity, etc.). False alarm probabilities were derived from insertion and retrieval experiments of simulated planets in the observed light curves. TEP was also the first thorough search for CBPs within the habitable zone of an MS-EB system, reaching a detection limit of super-Earth-sized planets. Detection probabilities of $\gtrsim$ 80% were achieved for any transiting CBP of 2.5 Earth radii or greater (Deeg et al. 1998, Doyle et al. 2000), after correlating over 400 million models with the more than 1000 hours of observations (see Fig. 6).

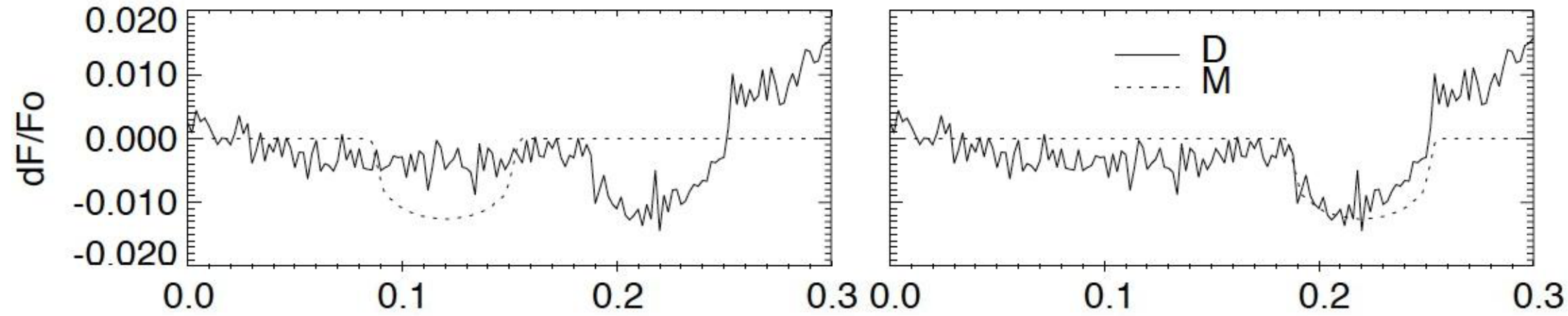

Fig. 6. A light curve from CM Draconis showing a possible planetary transit feature (solid line) along with the model light curve (dashed line) generated by the transit detection algorithm (TDA). In the left panel is a miss-fit of the orbital phase of the model while in the right panel is a good fit of the model against the possible transit signal (based on Doyle et al. 2000).

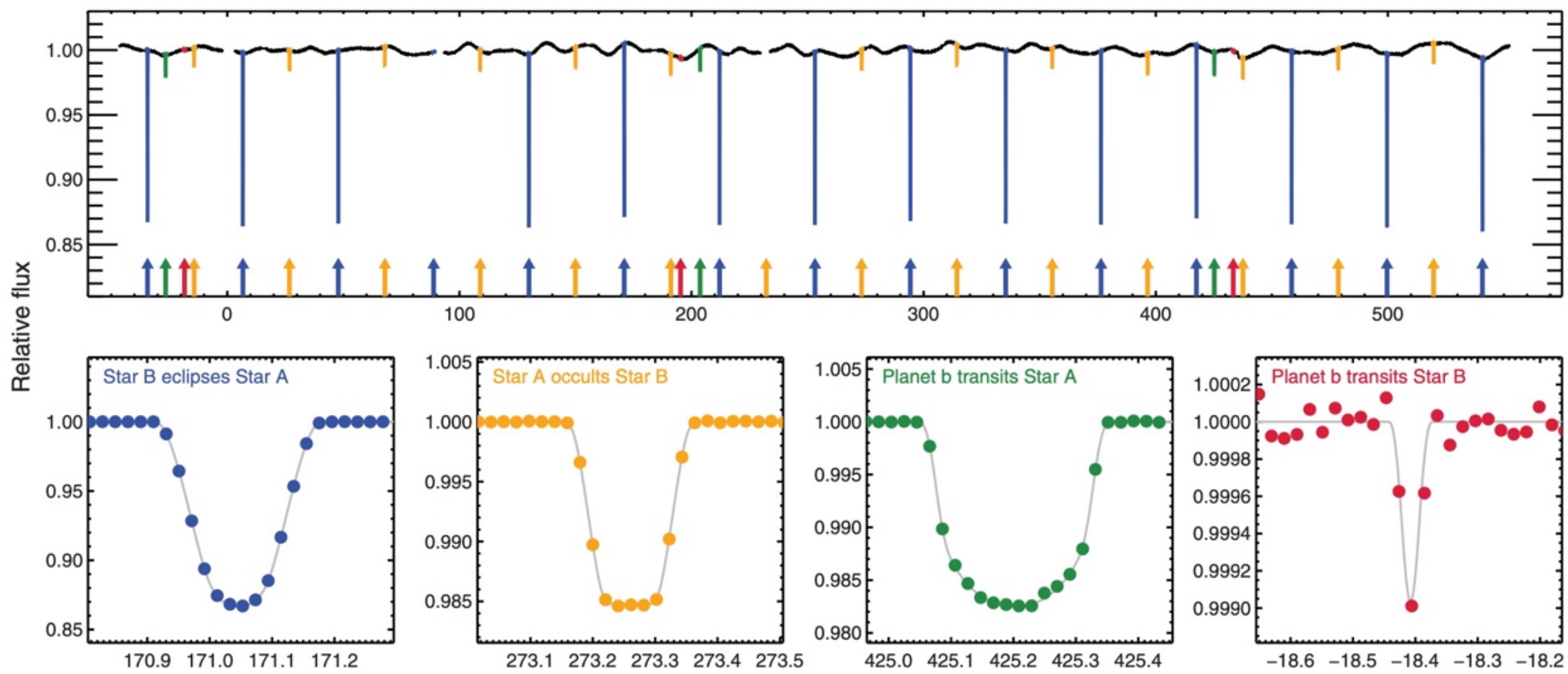

Fig. 7. Light curve of the Kepler-16AB b system – transits of the CBP across the brighter binary component A are in green, while CBP transits across the secondary component B are in red. The order of these transits can be seen (as indicated by the colored arrows at the base of the light curve) to reverse in the second pair of transits (red then green), compared to the first and last pair (green then red). Such a reversal is impossible to explain in terms of any other known astrophysical event (i.e., a background EB could not produce such a series of events in such a switching order) and so this allowed the confirmation that these events were caused by a circumbinary body. Reproduced from Doyle et al. (2011).

In 2011, the first CBP transits were discovered in data from the Kepler space mission (Doyle et al. 2011) around an EB then termed as Kepler 16. The very specific sequence of transits and binary eclipses (Fig. 7) confirmed the presence of a CBP, since such a sequence would not lend itself to any other astrophysical explanation. At present (April 2024), the transit method has provided the discovery of 14 CBPs, of which 12 were discovered in data from the Kepler mission and 2 in those from the TESS mission. CBP searches using the K2 (Kepler “second light) mission or the CoRoT mission data (Klagyivik et al. 2017) did not find any CBPs. These non-detections, and the sparsity of TESS detections, are due to these missions’ lower temporal coverage and their lesser photometric precisions. In the case of CoRoT, the longest coverages were of 5 months, versus the 4 years of Kepler, with similar number of targets surveyed (Klagyivik et al. 2017). In the case of TESS discoveries, one of its CBPs (TOI-1338 b,

Kostov et al. 2020) was found in the TESS continuous viewing zone (providing surveillance for one year), whereas the other CBP, TIC 172900988, was found in a single 28-day long TESS pointing, but required further observations (of further transits from ground and from space with the CHEOPS mission; and also eclipse timings and radial velocities) for confirmation (Kostov et al. 2021, Sairam et al. 2024). A more detailed review on space-based surveys for transiting CBPs, is given by Kostov (2023), and for more details on the CBPs detected by Kepler, see the chapter Two Suns in the Sky: The Kepler Circumbinary Planets.

In the distribution of the currently known transiting CBPs, two features are of note (Fig. 4): For one, most of them orbit near the inner stability limit around their central EB. This is likely caused by inward migration of the CBPs during the systems' evolution, with orbital resonances stopping them from further inward migration (Holman & Wiegert 1999, Pierens & Nelson 2013, Bromley & Kenyon 2015, Lai & Muñoz 2023). The other notable feature is a lack of CBPs around EBs with period of less than 7 days. CBP detections towards longer binary periods are expected to suffer strong detection biases (CBP orbits also have to be longer due to the inner stability limit; they are then less likely to be aligned correctly and if aligned, produce less transits per time); hence it was initially expected that most CBP transit detections would occur around the most common EBs with periods of a few days and less. For short orbit EBs, the tendency of CBPs to accumulate near the inner stability limit would even lend them to an alternative detection technique based on reflected light, with less constraints on orbital inclinations than posed by transits (see later in this chapter). Hence, CBPs around short-period EBs would be easier to detect, so that their absence is very likely a real feature in the distribution of CBPs. This absence has been explained by an orbital shrinking process that is required for the formation of close binaries, that may only proceed in the presence of a third stellar companion. This shrinking process would also be highly restrictive to the formation of detectable CBPs, either ejecting the planets or leaving them on large orbits on high mutual inclinations (Armstrong et al. 2014, Martin et al. 2015, Muñoz & Lai 2015, Hamers et al. 2016, Wenrui & Dong 2016, Kratter 2017, Moe & Kratter 2018). For these features in the transiting CBP distribution, and also a discussion of their typical sizes, with most in the Neptune to sub-Jupiter range (0.4 – 0.8 $R_{jup}$), see also the chapter Populations of Planets in Multiple Star Systems).

A more thorough exploration of the parameter-space of transiting CBPs is expected from the PLATO space mission, with a scheduled launch at the end of 2026 (Rauer et al. 2014, 2024, see also the chapter Space Missions for Exoplanet Science: PLATO). With observing durations that are similar to Kepler, but survey fields that are ~20 times larger, and a stellar sample that is ~1.5 mag brighter than Kepler's, Rauer et al. (2014) predicted that the number of known transiting CBPs becomes several times larger. Also, it is expected that CBPs discovered by PLATO will be more suitable for detailed characterizations than the current set of transiting CBPs, as more sensitive complementary observations, in particular those based on spectroscopy (e.g. radial velocities or transmission spectroscopy during transits) should become possible due to the brighter survey sample.

## Direct detection (Imaging)

Direct imaging involves capturing pictures of exoplanets by blocking out the overwhelming flux from the host stars. This method requires advanced instrumentation like coronagraphs or starshades, where instruments like the VLT/SPHERE (Very Large Telescope/Spectro-Polarimetric High-contrast Exoplanet Research) and the Gemini Planet Imager have been instrumental for progress, which can isolate the light from the planets against light from the central binary. In the context of circumbinary planets, direct imaging must account for the combined luminosity of two stars and -depending on the lateral distance between the stars – for the complexity of suppressing the light from two separate sources.

The advantages and disadvantages of direct detections by the imaging method are rather similar for the discovery and investigation of CBPs than for planets around single stars (see the chapter Direct Imaging as an Exoplanet Discovery Method). Direct detection is particularly effective for spotting planets that are far from their host stars, for which other methods like transits, radial velocities and astrometry loose efficiency. An advantage of planets that have been detected by imaging is the opportunity to measure directly their brightness and – at least in principle – to obtain their spectra separately from that of the central stars. To date, this method permitted only the detection of CBPs that are rather far away from their central stars – at distances on the order of tens to thousand AU – and with orbital periods in the range of 2000 to tens of thousands of years. All current imaging-detected CBPs are self-luminous objects in young systems, that still have elevated temperatures from their formation process. This makes them notable emitters of near-IR thermal radiation, thereby permitting their detection in that wave-band. For imaged CBPs, their masses can only roughly be estimated based on correlations between brightness, mass and age (e.g. Burrows et al. 1997, Mordasini et al. 2017), which furthermore depend on the planet formation history. A consequence of the imaged planets' relatively high brightnesses at a given age are however their rather high mass estimates. The six imaged CBPs that are currently listed in the NASA Exoplanet Archive have masses-estimates from 6 to over 20 $M_{jup}$, and only two of them (Ross 458 c with about 6$M_{jup}$ and ROXs 42 B b with 9±3 $M_{jup}$, based on Burgasser et al. 2010 and Currie et al. 2014, respectively) are below the canonical upper planetary mass-limit of 13 $M_{jup}$, while several others have a high likelihood to be really Brown Dwarfs.

The direct detection of non-luminous planets in reflected (instead of emitted) light is expected to become possible only with the operation of space based coronographic or interferometric imagers; for this, see the chapter Future Exoplanet Space Missions: Spectroscopy and Coronographic Imaging).

The relevance of imaging methods for the research on CBPs might not be that much from the detection of planets themselves, but rather from investigations of circumbinary disks as progenitors of planets. This involves not only imaging in the optical/IR domain, as shown in this chapter's introduction (Fig. 2), but more recently this has been extended to mm-wavelengths, thanks to the ALMA interferometer. As an example, the case of CS Chamaeleonis is given, which is a spectroscopic binary separated by ~4AU, with a T-Tauri component and an estimated age of 4.5 Myr. A known IR-excess led to prior suspicions for the presence of a disk, which became resolved in polarimetric NIR imaging with VLT/SPHERE (Ginski et al. 2018), showing a circumbinary disk with an outer radius of ~55 AU, but also a further companion at ~210 AU. That object might be a ~20 $M_{jup}$ brown dwarf or high-mass planet, potentially with its own disk or dust envelope. Later ALMA observations at 0.87 mm (Kurtovic et al. 2022) did not detect that companion, but led to a refinement of the binary disk morphology, finding an asymmetry and a slight eccentricity of 0.039. Based on disk evolution simulations, this suggests the existence of a Saturn-mass CBP close to the inner stability limit, at a distance of 3.3 – 4.0 times the binary separation.

## Gravitational Lensing (Microlensing)

Due to the warping of spacetime, as explained by general relativity, a star and its planets may temporarily (for hours to weeks, increasing with the mass of the lensing object) focus light from a background star onto Earth in a highly columnated way. The combined gravitational fields of binary stars create complex lensing patterns that may produce multiple caustic curves (Mao & Paczynski, 1991). A CBP may cause additional perturbations in these curves, leading to distinctive signatures that may permit its identification. Distances to the lensing system are usually significant (kiloparsecs) and so CBPs can be sampled at great distances. Unlike the transit or radial velocity methods, which are more sensitive to planets in close orbits, microlensing is more likely to detect planets at wider separations from their host binary. Neither is this method affected by the stellar activity of the central stars, which makes it a robust method for studying a variety of stellar environments.

The first microlensing detection of a circumbinary planet was reported by the Optical Gravitational Lensing Experiment (OGLE) collaboration in the system OGLE-2007-BLG-349 (Bennett et al. 2016). After follow-up observations with the Hubble Space Telescope that resolved its stellar-neighborhood, the interpretation as a stellar binary with a further CBP provided the only consistent model for the observed light curve. The system is at a distance of about 2.76kpc and the CBP has a mass $m_c = 80 \pm 13\ m_\oplus$, orbiting a pair of M-dwarfs with masses of $m_A = 0.41 \pm 0.07\ m_\odot$ and $m_B = 0.30 \pm 0.07\ m_\odot$. The binary has a period of about 9 days, while the planet is at an approximate projected (lateral) distance of 2.6AU, implying an orbital period of about 7 yr. It is of note that the binary period is within the regime covered by the transit method, while the CBP's period is about an order of magnitude larger than that of the transiting CBPs.

All CBPs detected by microlensing orbit on distances from the barycenter of 2.6 – 6.4 AU, with orbital periods in the several years to several decades-long range. Similar to microlensing detections of planets around single stars, the distances among the lensing objects, and their masses, remain rather uncertain; for the orbital distances, the quoted errors range from 25 - 90% of the values themselves. Also, only the lateral distances between a CBP and the central binary are estimable from microlensing, so their real distances, and hence the planet's orbital periods, might be significantly larger, and designations as CBP systems might not be secure ones. As an example, the case of OGLE-2023-BLG-0836L is quoted (Han et al. 2024), in which the most likely lateral separation between the binary components is 1.9 AU, whereas the separation between the primary and the planet-like object (of 4.4 $M_{jup}$) is 3.7AU. With the distance to the planet being only twice the binary-separation, this would be an unstable configuration; therefore, the real separations between the three objects have to be rather different. As an alternative, Han et al. indicate that a planet orbiting only one of the binary components cannot be entirely ruled out as well.

## Eclipse Timing Variation from Dynamical Interactions

A CBPs that is sufficiently close to its central binary may cause detectable perturbations to the binary's Keplerian orbit. Such binary orbits may then become significantly different from those that are representable by 2-body systems, in which the binary orbit is determined only by the binary. It is of note that such dynamical effects due to the CBP are different to the LTTE introduced previously, in which the binary orbit can be treated as a 2-body problem, with the binary's barycenter orbiting around the entire system's (including one or more CBPs) barycenter.

If the orbital period of a CBP is relatively short (of years or less), then the effects of the LTTE will become too small to be detectable by ETVs. However, the closer a CBP is to the EB, the more gravitational pull does it exert on the stellar components directly, leading to orbital perturbations that may cause detectable deviations from Keplerian orbits. That is, the binary orbits will become significantly different to those from 2-body systems, in which the binary orbit is determined by the binary.

Dynamical effects from an orbiting CBP may occur in three classes of orbital perturbations (Brown 1936). They are: short-period perturbations on time-scales of the inner binary period ($P_{bin}$); long-period perturbations on scales of the outer orbital period ($P_3$), and apse-node perturbations on still longer scales of $P_3^2 / P_{bin}$. The primary observable for dynamical effects are eclipse timings that indicate variations in the binary period, but other observables may arise as well: variations in the times (or phases) of secondary eclipses relative to primary eclipses may indicate changes in the angle of periastron and potentially in eccentricity, and variations in eclipse depth and duration may indicate changes in orbital inclination.

For perturbations on time-scales of $P_3$ – which are typically most accesible to observations (given that amplitudes of the short-period perturbations tend to be very small, while $P_3^2 / P_{bin}$ tends to be a very long time) – the amplitude of the offset of eclipse times versus strict periodicity (O-C) can be characterized by (Borkovits et al. 2003):

$$K_{O-C} = \frac{3\, m_3}{8\pi\,(m_{bin} + m_3)} \frac{P_{bin}^2}{P_3} \qquad (3)$$

It is of note that the amplitude increases as the period of the CBP gets shorter, in the opposite direction of Eq. 1 for the LTTE. Also, since gravitational dynamical effects are independent of the inclination against the observer (without a *sin i* dependency, as in Eq. 1), the direct mass of a CBP can in principle be derived from observed O-C values. While above discussion refers to timing variations from longer-term orbital evolution, shorter-term timing variations on time-scales of the orbital periods may also occur; see Borkovits et al. (2011) and references therein.

The usefulness of interpreting eclipse timings as dynamical effects was demonstrated during the validation and characterization of the first CBP detected by transits, Kepler-16b (Doyle et al. 2011). Examining its binary eclipse timings against the LTTE, it was only possible to constrain the mass of the third body to be less than that of a T-type brown dwarf. An interpretation of the same timings as dynamical effects due to the third body – especially the drift in the timing of the secondary eclipses – was however able to tightly confirm the mass of Kepler-16b to one of 0.333±0.016 $M_{jup}$ (Fig. 8). The LTTE timing amplitude corresponding to that mass is only 0.13 sec; hence that LTTE is undetectable in the Kepler-data. Similar detections of the dynamical effect have later been reported for several more CBPs detected by transits (e.g. Kepler 34 and 35, Welsh et al. 2012).

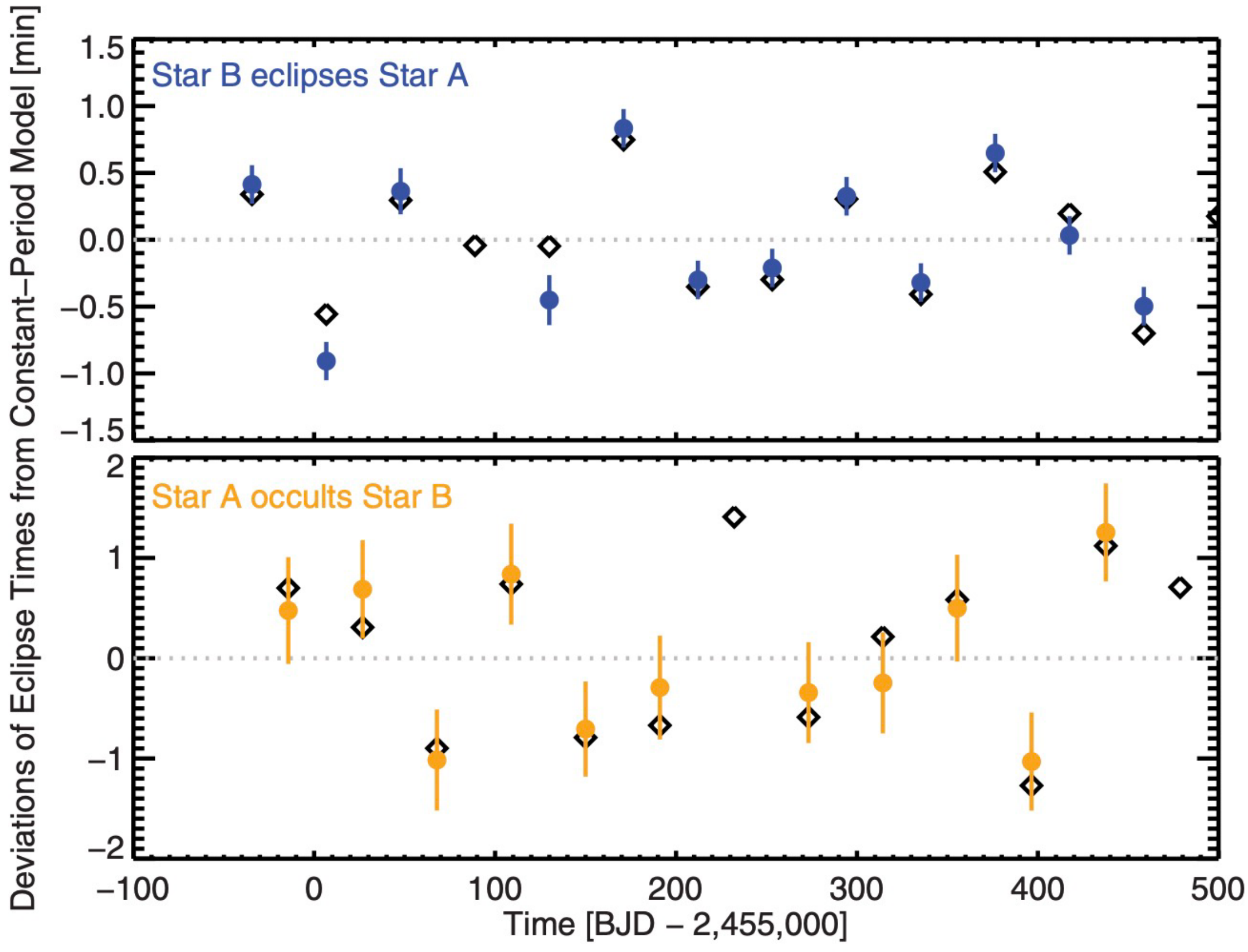

Fig. 8. Deviations of Kepler-16 eclipse times of primary (top panel) and secondary binary eclipses (bottom panel) from strict periodicity, in light curves of the first 540 days from the Kepler mission. The interpretation of these deviations as dynamical effects caused by the planet implied a planetary mass of 0.333±0.016 $M_{jup}$. Reproduced from Doyle et al. (2011).

Currently, some observational efforts are going on to detect ETVs from both the LTTE and dynamical effects; notable is the SOLARIS photometric survey that concentrates on binaries with periods of 0.7 – 2 d, with the specific aim to detect circumbinary companions (Moharana et al. 2024). To date, no CBP discoveries have been secured solely based on ETV interpretations from dynamical effects. However, dynamical-effect interpretations of ETVs have led to CBP candidates that a posteriori were confirmed from radial velocities; such cases are presented in the next section on detections by radial velocities.

## Radial Velocity Variations

In contrast to the detection of planets around single stars, where radial velocity (RV) measurements are one of the dominant detection methods (see chapter Radial Velocities as an Exoplanet Discovery Method), RVs have played a lesser role in the discovery of CBPs. This is due to the difficulties in detecting RV variations caused by a CBP in the presence of the much larger RV variations from the stellar motion within the binary.

In a two-body system with masses $M_A$ and $M_B$, the RV amplitude $K_{A_B}$ of component A due to component B is given by:

$$K_{A_B} = \left(\frac{2\pi G}{P_{AB}}\right)^{1/3} \frac{M_B \sin i}{(M_A+M_B)^{2/3}} \left(1-e^2\right)^{-1/2} \ , \qquad (4)$$

where $P_{AB}$ , $i$, $e$ are the system's orbital period, inclination and eccentricity. If a binary AB is orbited by a CBP of identical inclination and eccentricity and with a mass $M_p$ << ($M_A$ + $M_B$), the ratio of RV amplitudes due to the orbiting CBP and due to the stellar orbits is given by:

$$\frac{K_p}{K_{A_B}} = \left(\frac{P_p}{P_{AB}}\right)^{-1/3} \frac{M_p}{M_B}, \qquad (5)$$

where $K_p$ is the RV amplitude of either of the stellar components due to the presence of the planet. The dominant factor in Eq. 5 is clearly the mass ratio. For a binary consisting of solar-mass components, orbited by a Jupiter-mass CBP with a period ratio of 3.5 (near the inner stability limit), the ratio of RV amplitudes is about 1/1600; becoming even lower for wider-orbit CBPs. Even for M-star binaries orbited by massive CBPs of several jovian masses, this ratio would not exceed about 1/100. The extraction of a CBP's RV signal, of typically 1–- 100 m/s, is therefore very difficult in the presence of the much larger signal from the central binary, with typical RV amplitudes of 10 - 100 km/s.

The first potential CBP discovery by RVs was brought forward by Correia et al. (2005) in a paper entitled "A pair of planets around HD 202206 or a circumbinary planet?". They describe a system with these components: A solar like G6V central star, a *b* component with a mass of *m* sin *i* ~17.4$M_{jup}$ and a *c* component with a mass of *m* sin *i* ~2.44$M_{jup}$ , with both *b* and *c* on markedly eccentric orbits. Acceptance of *c* as a CBP or as a second planet depends on the component *b* being considered as a brown dwarf or a planet (see chapter Definition of Exoplanets and Brown Dwarfs). This system also presents serious issues on its long-term stability, which favors a 5:1 mean motion resonance between the two orbiting bodies (Correia et al. 2005, Couetdic et al. 2010). However, Benedict & Harrison (2017) showed from astrometric parallaxes that HD 202206 is a nearly face-on system, with small values of sin *i* and consequently, much larger orbital masses for *b* and *c*, of about 94 $M_{jup}$ and 18$M_{jup}$. Hence, they conclude that none of the alternatives proposed by Correia et al. (2005) apply; HD 202206 is instead a G8V + M6V binary orbited by a brown dwarf. (However, in the NASA exoplanet archive it continues to be listed as a CBP system discovered by RVs.)

An ambitious CBP search effort by RVs was undertaken with the TATOOINE search by Konacki et al. (2009). Starting in 2003, several spectrographs with iodine absorption cells (such as Keck/HIRES or the TNG/Sarg instrument) were used to survey a set of double-lined (SB2) binaries, using the TODCOR (Zucker & Mazeh 1994) algorithm to disentangle the RVs. However, with a reported precision of 7 – 15 m/s, no CBP has been announced by that project.

BEBOP ("Binaries Escorted By Orbiting Planets", Martin et al 2019, Standing et al. 2022) was initiated in 2013 as a survey of single lined (SB1) binaries with highly unequal masses. This selection provides preferential biases in RV amplitudes and avoids spectral contamination due to overlaps of the components' spectra. BEBOP began with the Coralie spectrograph on the 1,2m Euler telescope in Chile and reached sensitivities to CBPs of masses ≳0.5 Mjup on several-year-long orbits (Martin 2020). This permitted it to demonstrate the absence of a sizable population of more massive CBPs on large mutual inclinations (against the binary plane), previously proposed by Martin & Triaud (2014). The survey was later expanded to include HARPS on the 3,6m ESO telescope and SOPHIE on the 1.93m OHP in France, and has reached precisions of 3m/s, becoming sensitive to Neptune/Saturn mass planets with periods up to 1000d (Standing et al. 2022). BEBOP was able to detect the RV signature of Kepler 16 b (Triaud et al. 2022) – the first CBP discovered by transits – and to provide its mass. It also found the first unambiguous CBP discovered by RVs, in the form of TOI 1338 c / BEBOP-1 c (Standing et al. 2023), which is the second CBP in the TOI-1338 system. The first CBP in TOI 1338 had been discovered by transits (Kostov et al. 2020); the RVs in combination with eclipse timings from TESS were later used by Wang & Liu (2024) to substantially refine the parameters of all components of TOI 1338.

RV observations have also become relevant as a means to refine and confirm CBP candidates found by other means. TIC 172900988 b was initially proposed as a CBP by Kostov et al. (2021) due to ETVs indicative of a third body and from a single transit from TESS. From an RV survey of double-lined binaries with SOPHIE (including several from the original TATOOINE sample) and with improved analysis methods based on Gaussian process regression, Sairam et al. (2024) were able to confirm this object as a CBP. Another case of RV observations being essential in the confirmation of a CBP was KIC 5095269 b / Kepler 1660 ABb, initially identified by Borkovits et al. (2016) due to ETVs in Kepler data, with Getley et al. (2017) proposing it as a CBP on a highly inclined orbit. Later RV observations (Goldberg et al. 2023) did not detect the CBP directly; they permitted however a substantial refinement of the binary's physical and orbital parameters. This, in turn, enabled an improved interpretation of the binary's eclipse timing and depth variations, thereby deriving reliable parameters of the third object and confirming it as a CBP, but now on a coplanar orbit.

## Astrometry

The detection of CBPs by astrometry in data from the GAIA mission was evaluated by Sahlmann et al. (2015), who predicted that GAIA astrometry has the potential to detect hundreds of giant CBPs on periods of a few years, while providing also direct measurements of their mutual inclinations. Currently, the only report about an astrometric detection of a CBP is on the common-envelope eclipsing binary HW Vir, previously reported with one or more CBP candidates found by eclipse timing, some of them in conflict with each other (Baycroft et al. 2023 and refences therein). Baycroft et al. combined GAIA DR3 and Hipparchos astrometric data and provided slight evidence for a proper motion anomaly due to a circumbinary companion. These data permitted them to place an upper mass-limit for such a companion, and excluded some of the previously claimed CBP configurations. Future GAIA data releases should lead to a clearly detectable proper motion signal if there is indeed one present, and to the discovery or confirmation of further CBPs.

## Reflected Light or Eclipse Echoes

The eclipse echo method proposes to observe the light of an EB that is reflected by a CBP and to detect the binary eclipses in that reflected light (Deeg and Doyle 2011). Besides the confirmation as a CBP, it would allow the derivation of several of the CBP's orbital parameters. Since a CBP 'sees' the eclipses of its central binary at another time than an external observer, the 'eclipse echoes' in the planet's reflected light would appear at different times and with a slightly different 'echo period', $P_{re}$, than directly observed eclipses, given by

$$1/P_{re} = 1/P_{AB} \pm 1/P_p \quad , \qquad (6)$$

where the ± sign indicates prograde (-) or retrograde (+) planetary orbits (Fig. 9). Considering a minimum in the planet/binary period ratio of about 3.7, $P_{re}$ gets constrained to values fairly close to the binary period, namely: $0.78\ P_{AB} \leq P_{re} \leq 1.37\ P_{AB}$. If only prograde orbits are considered, then the echo-periods are still more constrained, to $P_{AB} < P_{re} \leq 1.37\ P_{AB}$. This restricted range of periods should facilitate any echo detection attempts. Furthermore, the times of the echoes are also subject to minor (likely undetectable) variations due to dynamical interactions among the three bodies or due to the LTTE.

The eclipse echo method is most sensitive to short-periodic CBPs around binaries that are close to the inner stability limit. However, in principle this method could detect binaries and CBPs over a wide range of inclinations and – if successful – the orbital characteristics of the CBP system could be obtained solely from the photometric light curve.

After the development of this method, it became however evident that close CBPs around binaries with periods of less than 7 days are rare or even absent (e.g. Martin et al. 2015); systematic CBP searches with this method were therefore never performed. If such short-periodic CBPs would be detected at some point, the eclipse echo method could however become useful for a more complete characterization of such a planet, in particular to determine a planet’s mutual inclination and its albedo.

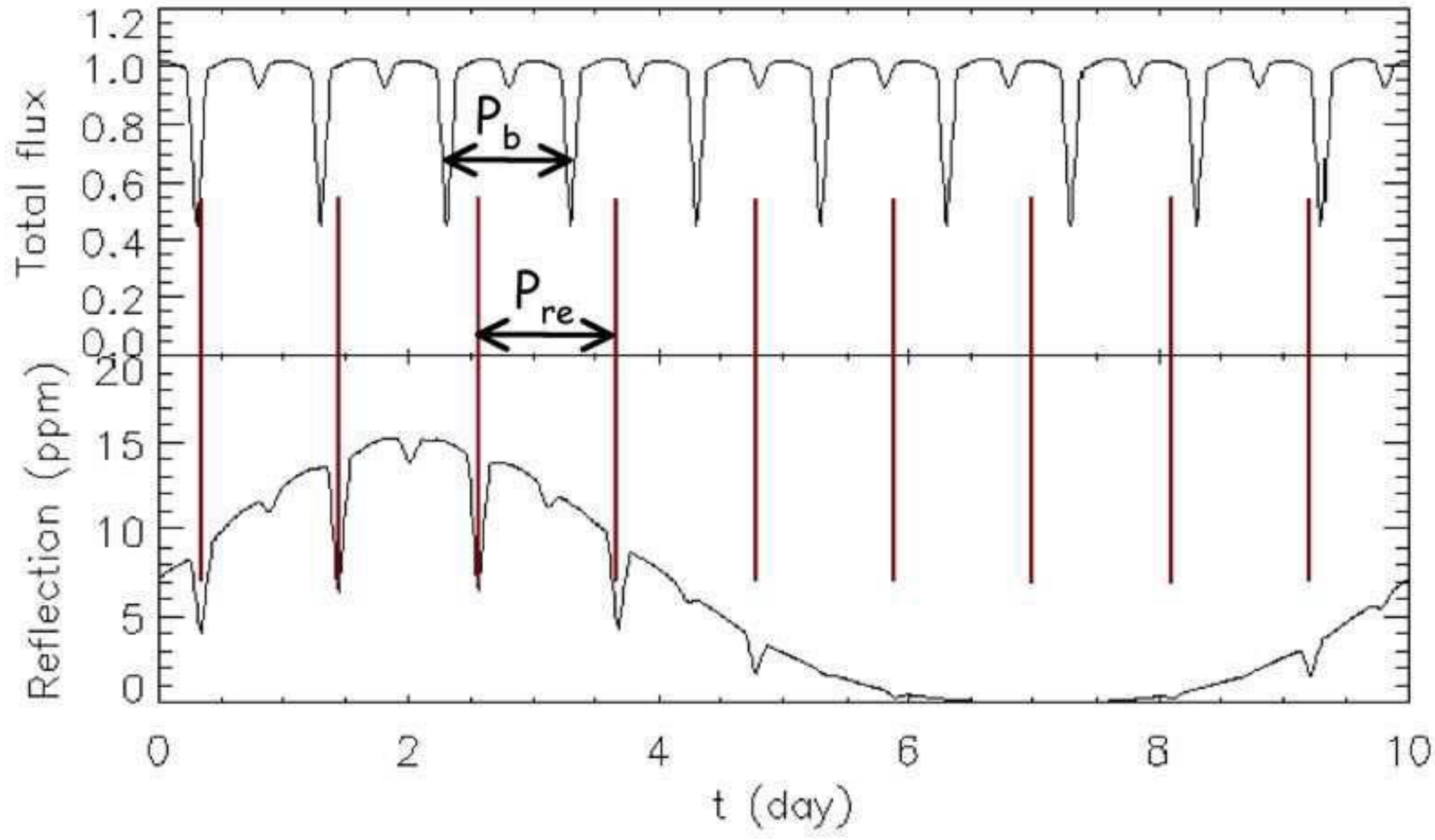

Fig. 9. Simulated light curve from a system with a binary of 1-day orbital period ($P_b$), orbited by a prograde planet with a period of $P_p$ = 10 d, both with an orbital inclination of 90 degrees. The upper panel shows the total observed flux, which is dominated by the eclipses observed directly from the binary. The lower panel shows the reflected light from the planet. Note that its amplitude is on the order of ppm. The general shape of the reflected light curve is due largely to the planetary phase function, upon which the eclipse echos (positions indicated by vertical red lines) are imprinted, with a periodicity of $P_{re}$ = 1.11 d (From Deeg & Doyle 2011).

## Habitability of Circumbinary Planets

With the beginning of the first transit searches for CBPs, interest in the habitability of CBPs arose (Fig. 10). Habitable zones (HZs) around CBPs differ from those around single stars mainly by the fact that a CBP receives a quite variable stellar flux due to the rapidly changing distances between the stellar components and the planet, as well as the regular dimming due to stellar eclipses, at least for CBPs in or near the EB orbital plane (see Welsh et al. 2012 for a calculation of the changing insolation reaching Kepler 34b and 35b). These features definitively have an impact on habitability (Haghighipour & Kaltenegger 2013, Popp & Eggl 2017, Kong et al. 2022), but not necessarily in a negative sense. CBPs may be very resilient against circumbinary-driven climate variations (Wolf et al. 2020; see also Mason et al. 2015); it has even been argued that CBPs around MS-binaries are particularly predisposed for habitable conditions (Shevchenko 2017). Further recent discussions on CBP habitability are given by Georgakarakos (2021) for the CBPs found by TESS; and by Georgakarakos et al. (2021) for the CBP – HZ in the presence of another giant planet. For a more detailed evaluation of the habitability of CBPs, see the chapter Habitability of Planets in Binary Star Systems.

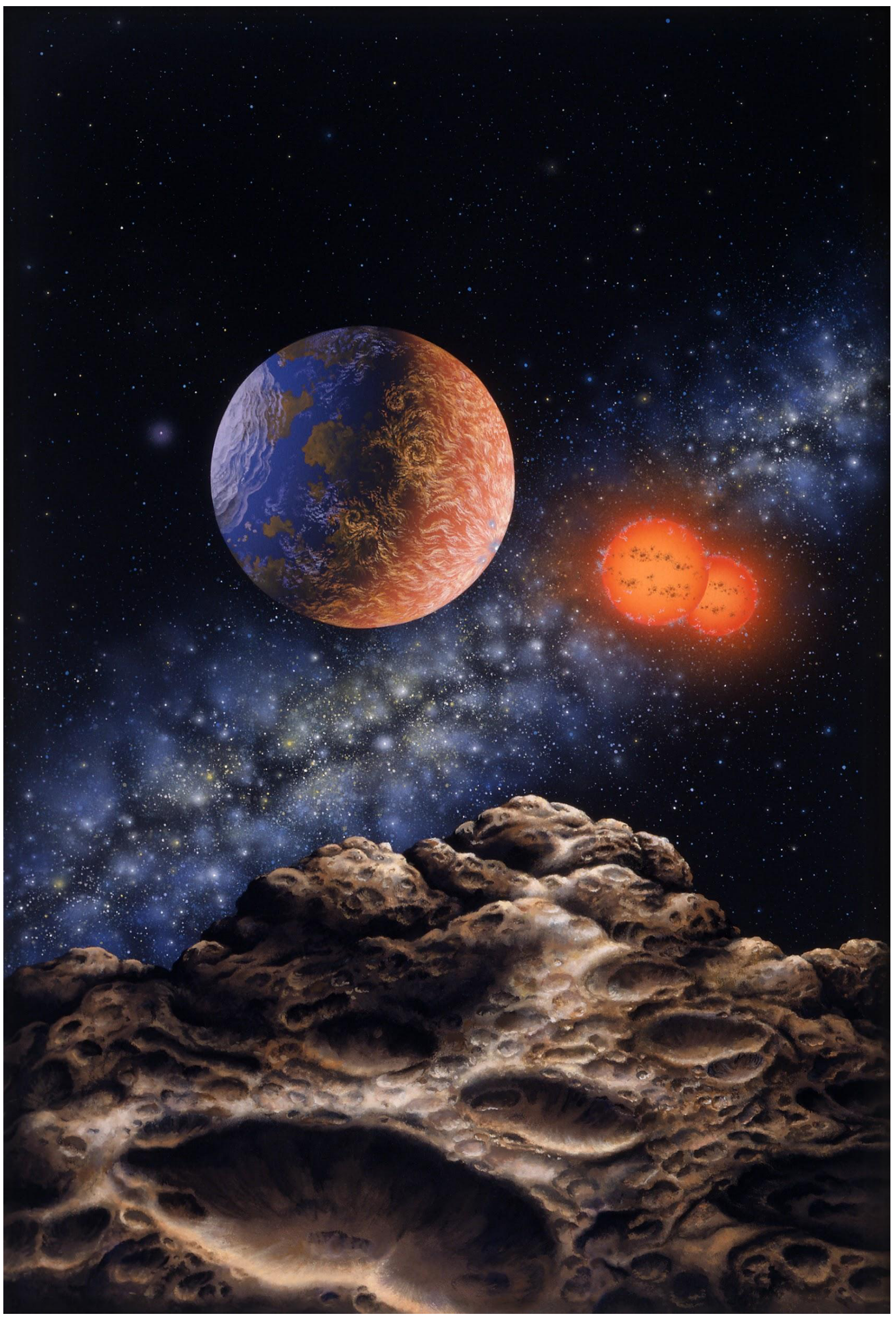

Fig. 10. Artist's conception (Lynette Cook, reproduced with her permission) of a tidally-locked terrestrial planet orbiting within the HZ around the CM Draconis eclipsing binary system. The weather patterns propagate not across lines of latitude, but in streams from the substellar point (Heat et al. 1999), and there is an equatorial ice cap at the anti-substellar point, which is illumined by a white dwarf proper motion companion (brighter star above the ice cap). CM Draconis was the first binary star system searched for CBPs within its habitable zone.

# Conclusions

For various reasons there were doubts about the existence of CBPs until the discovery of Kepler-16b in 2011. This system was nicknamed "Tatooine" after one of the planets in the Star Wars movie because – like the sunset scene in the movie – it was the first CBP system where the binary pair could be seen as approximately solar-sized discs in the sky from the perspective of the planet. (George Lucas, the producer of Star Wars, gave unofficial permission to use the nickname "Tatooine" for this system, and the name stuck.)

The CBPs detected to date by various methods are shown in Fig. 4. CBPs whose confirmation depended on several detection methods, like Kepler 1660AB b or TIC 172900988, have recently become more common. For them, a combination of transits, eclipse timings and RVs were needed for the establishment of basic parameters that constitute a reliable planet discovery.

From the current set of known CBPs, it can be estimated that there must be at least hundreds of millions of such planetary systems in our Galaxy (for a more detailed discussion, see the chapter Populations of planets in binary star systems). CBPs are among the most precisely measured planetary systems available, giving insights into star-planet spectral type and size distributions, and putting constraints on star and planet formation. Several questions on MS-CBPs have been around since the discovery of their first batch in data from the Kepler mission, which have not been resolved fully to date: Why are many CBPs near their inner orbital stability limit; why do CBPs have intermediate masses (i.e., Neptune to Saturn masses; see Armstrong et al. 2014); and why have no CBPs been found around shorter-periodic (< 7 days) MS binaries? The stability of 'misaligned' CBPs with orbits that are strongly inclined against the binary orbit has been demonstrated (Martin & Triaud 2014, 2016) but their existence, for which several detection methods such as eclipse timing variations (Borkovits et al. 2011) or occasional transits (Martin 2017) are suitable, is yet to be demonstrated. Some detection methods – in particular transits, radial velocities and eclipse timing due to dynamical effects – cover similar parameter spaces for CBPs and central binaries and in some systems, they are able to complement each other. On the other hand, CBPs found by the LTTE form a well-distinguished group against the former one, while those from microlensing are not that well separated from it. This leads to the question if there are continuous CBP populations between planets detected by different methods. If there is no continuity, it is to be established if the absence of current detections means a true absence of such CBPs (for example, around binaries with P ≈ 1d) or if absences might have arisen from biases in the detections. In a historic context, the study of CBPs has only just begun, while in the near future, it is expected that upcoming missions like PLATO will greatly improve our understanding of these strange worlds.

# Cross-References

- PSR B1257+12 and the First Confirmed Planets Beyond the Solar System
- Two Suns in the Sky: The Kepler Circumbinary Planets
- Circumbinary Planets Around Evolved Stars
- Populations of Planets in Binary Star Systems
- Habitability of Planets in Binary Star Systems
- Transit Photometry as an Exoplanet Discovery Method
- Direct Imaging as an Exoplanet Discovery Method
- Radial Velocities as an Exoplanet Discovery Method
- Transit-Timing and Duration Variations for the Discovery and Characterization of Exoplanets
- Space Missions for Exoplanet Science: Kepler/K2
- Space Missions for Exoplanet Science: PLATO
- Space Missions for Exoplanet Science: TESS
- Microlensing Surveys for Exoplanet Research (OGLE Survey Perspective)

# Acknowledgements

HD acknowledges support by the Spanish Research Agency of the Ministry of Science and Innovation (AEI-MICINN) under the grant PID2019-107061GB-C66. This contribution has benefited from the use of the NASA Exoplanet Archive and the Extrasolar Planets Encyclopaedia and the authors acknowledge the people behind these tools.